\documentclass{PoS}

\usepackage{amsfonts,amsmath,graphicx,bm,siunitx}
\usepackage{dcolumn}
\usepackage{ifpdf}
\usepackage{multirow}
\usepackage{float}

\title{$|V_{cb}|$ from the $\bar{B}^0\rightarrow D^{*+} \ell^- \bar{\nu} $ zero-recoil form factor using 2$+$1$+$1 flavour HISQ and NRQCD}

\ShortTitle{$|V_{cb}|$ from $\bar{B}^0\rightarrow D^{*+} \ell^- \bar{\nu} $ using NRQCD-HISQ}

\author{\speaker{Judd Harrison}%
       \\
       Department of Applied Mathematics and Theoretical Physics, University of Cambridge, Cambridge, CB3 0WA, UK \\
       E-mail: \email{jgihh2@cam.ac.uk}}

\author{Christine Davies%
       \\
SUPA, School of Physics and Astronomy, University of Glasgow, Glasgow G12 8QQ, UK\\
E-mail: \email{Christine.Davies@glasgow.ac.uk}}

\author{Matthew Wingate%
       \\
       Department of Applied Mathematics and Theoretical Physics, University of Cambridge, Cambridge, CB3 0WA, UK \\
       E-mail: \email{M.Wingate@damtp.cam.ac.uk}}

\author{HPQCD Collaboration}

\abstract{We present the status of our ongoing calculation of the zero-recoil form factor for the semileptonic decay $\bar{B}^0\rightarrow D^{*+}l^-\bar{\nu}$ using lattice QCD with 2+1+1 flavours of highly improved staggered quarks in the sea (the MILC HISQ configurations) and using non-relativistic QCD for the bottom quark. We combine our result for $ F(1)$ with the latest HFAG average of $\eta_{EW} F(1)|V_{cb}|$ to get a preliminary value for $|V_{cb}|$.}

\FullConference{34th annual International Symposium on Lattice Field Theory\\
	 24-30 July 2016\\
	 University of Southampton, UK}

\begin{document}

\section{Introduction}

Tests of the unitarity of the Cabibbo-Kobayashi-Maskawa (CKM) matrix form a key check of the consistency of the standard model of particle physics. The unitarity conditions require that the elements of the CKM matrix satisfy $\sum_j V_{ij}V^*_{kj} = \delta_{ik}$ where, following the usual notation, $i$ and $k$ take values $u,c,t$ and $j$ takes values $d,s,b$. Currently the elements involved in transitions between third generation quarks and the lighter quarks are least well determined~\cite{REVIEW}. Of these elements, all with roughly commensurate errors, we focus on the exclusive determination of $|V_{cb}|$ using the semileptonic $B\rightarrow D^*$ decay. The 2014 edition of the Review of Particle Physics \cite{REVIEW} by the Particle Data Group notes that the value of $V_{cb}$ determined via exclusive decays of the $B$ meson to either a $D$ or $D^*$ and the inclusive determination, using all $B$ decays involving a $b\rightarrow c$ transition, are only marginally consistent. The most recent determination from lattice QCD using the Fermilab action for both $c$ and $b$ quarks reveals some tension \cite{FERMILABDSTAR}. We have therefore undertaken here a new calculation which differs significantly in approach. Our calculation uses a relativistic HISQ valence $c$ quark, a perturbatively improved nonrelativistic action for the $b$ quark and improved gauge field configurations \cite{MILC1,MILC2} that include highly improved staggered quarks (HISQ) in the sea with reduced taste-exchange violations \cite{HISQ} and physical light, strange and charm quark masses. 

The decay rate can be parameterised as
\begin{equation}
\frac{d\Gamma}{d\omega}(\bar{B}^0\rightarrow D^{*+}l^-\bar{\nu}_l) = (1+\pi \alpha)\frac{G_F^2M_{D^*}^3}{4\pi^3} (M_B-M_{D^*})^2(\omega^2-1)^{1/2}\chi(\omega)|V_{cb}|^2|\eta_{EW} F(\omega)|^2\nonumber
\end{equation}
where $\omega = v\cdot v'$ is the product of the $B$ and $D^*$ four-velocities, $\chi(\omega)$ is a phase space factor, $\eta_{EW}$ accounts for electroweak corrections due to box diagrams in which a photon or $Z$ boson is exchanged in addition to a $W$ boson, $1+\pi\alpha$ accounts for the Coulomb attraction of the final-state charged particles and $ F(\omega)$ is the form factor coming from QCD. Many experiments have measured the differential rate as a function of $\omega$. One output of fits to such data is $\eta_{EW}  F(1) |V_{cb}|= 35.81(11)(44)\times10^{-3}$ \cite{HFAG}. $\eta_{EW}$ can be estimated perturbatively with an error at least an order of magnitude smaller than we are concerned with ~\cite{ETAEW}. All that is then needed for an accurate determination of $V_{cb}$ is the value of $ F(1)$. Current lattice predictions give $ F(1) = 0.902\pm 0.017$ \cite{REVIEW} which, combined with the experimental result and value for $\eta_{EW}$, yields $|V_{cb}| = 39.48 (50)_{\text{exp}} (74)_{\text{theory}}\times 10^{-3}$, in tension with the inclusive result $|V_{cb}| = 42.2 (0.7)\times 10^{-3}$.

\section{Lattice Setup}
The gluon field configurations that we use were generated by the MILC collaboration and include 2+1+1 flavours of dynamical HISQ quarks in the sea. The $u$ and $d$ quarks have equal mass, $m_u = m_d \equiv m_l$, and we use the values $m_l/m_s = 0.2$, $0.1$ and the physical value $1/27.4$ \cite{lightmassratio}. The $s$ and $c$ quarks in the sea are also well-tuned \cite{PhysMasses} and included using the HISQ action. The gauge action is the Symanzik improved gluon action \cite{GaugeAction} with coefficients correct to $\mathcal{O}(\alpha_s a^2, n_f \alpha_s a^2)$. The $b$ quark is simulated using non-relativistic QCD \cite{NRQCD,latticespacing}, which takes advantage of the non-relativistic nature of the $b$ quark dynamics in $B$ mesons and produces very good control over discretisation uncertainties.

\begin{table}[H]
\centering
 \begin{tabular}{c c c c c c c c c} 
 \hline\hline
Set &  $a(\text{fm})$ & $L/a \times T/a$ & $am_l$ & $am_s$ & $am_c$  & $\alpha_V(2/a)$ & $u_0$ & $n_\text{cfg} \times n_\text{t}$\\ [0.1ex] 
 \hline
1 & 0.1474 & $16 \times 48$ & 0.013 & 0.065 &0.838   & 0.346 & 0.8195  & 960$\times$16 \\ 
2 &  0.1463 & $24 \times 48$  & 0.0064  & 0.064&0.828  & 0.344 & 0.8202 &480$\times$4\\
3 & 0.145 & $32 \times 48$  & 0.00235  & 0.0647&0.831 & 0.343 & 0.8195  &960$\times$4\\
 \hline
4 & 0.1219 & $24 \times 64$  & 0.0102 & 0.0509&0.635   & 0.307 & 0.8341 &720$\times$4\\
5 & 0.1195 & $32 \times 64$  & 0.00507  & 0.0507&0.628 &  0.308 & 0.8349  &960$\times$4\\
6 & 0.1189 & $48 \times 64$  & 0.00184 & 0.0507&0.628  & 0.307 & 0.8341 &720$\times$4\\

 \hline
7 & 0.0884 & $32 \times 96$  & 0.0074  &0.037&0.440 &  0.267 & 0.8525 &960$\times$4\\ 
8 & 0.08787 & $64 \times 96$  & 0.00120  & 0.0363 & 0.432  & 0.267 & 0.8518 &540$\times$4\\ 
 \hline
\end{tabular}
\caption{Details of the gauge configurations used in this work. We give the values for $\alpha_V(2/a)$ used in the matching relations. We refer to sets 1, 2 and 3 as `very coarse', sets 4, 5 and 6 as `coarse' and sets 7 and 8 as `fine'. The lattice spacings were determined from the $\Upsilon (2S - 1S)$ splitting in \cite{latticespacing}.  Sets 3, 6 and 8 use light quarks with their physical masses. $u_0$ is the tadpole improvement factor, here we use the Landau gauge mean link. The final column specifies the total number of configurations multiplied by the number of different start times used for sources on each. In order to improve statistics we use random wall sources. }
\label{tab:gparams}
\end{table}

\begin{table}[H]
\centering
 \begin{tabular}{c c c c c c c c c} 
 \hline\hline
Set &   $am_s^\text{val}$ & $am_c^\text{val}$ & $am_b$ & $\epsilon_\text{Naik}$& $c_1$,$c_6$ & $c_5$&$c_4$ & T \\ [0.1ex] 
 \hline
1 & 0.0641& 0.826 & 3.297& $-0.345$ &1.36 &1.21 &1.22& 10,11,12,13  \\ 
2 & 0.0636& 0.828 & 3.263&$ -0.340$  &1.36 &1.21 &1.22& 10,11,12,13 \\
3 & 0.0628& 0.827 & 3.25&$ -0.345$ & 1.36& 1.21&1.22& 10,11,12,13 \\
 \hline
4 & 0.0522& 0.645& 2.66& $-0.226$ &1.31 &1.16 &1.20 & 10,11,12,13\\
5 & 0.0505& 0.627 &2.62& $-0.224$ &1.31 & 1.16&1.20& 10,11,12,13\\
6 & 0.0507 & 0.631  &2.62 & $-0.226$ &1.31 & 1.16&1.20& 10,11,12,13\\
 \hline
7 & 0.0364&0.434 &1.91&$ -0.117$ & 1.21&1.12 & 1.16& 15,18,21,24\\
8 & 0.0360 & 0.4305 &1.89&$ -0.115$ &1.21 &1.12 & 1.16& 10,13,16,19\\ 
 \hline
\end{tabular}
\caption{Valence quark masses and parameters used to calculate propagators. The $s$ and $c$ valence masses are taken from \cite{PhysMasses} and the $b$ mass from \cite{latticespacing}. $(1+\epsilon_\text{Naik})$ is the coefficient of the charm Naik term and $c_{i}$ are the perturbatively improved coefficients appearing in the NRQCD action correct through $\mathcal{O}(\alpha_s v^4)$  \cite{latticespacing}. The last column gives the T values used in three point functions.}
\label{tab:vqparams}
\end{table}

In the case of the pseudoscalar to vector decay the only contribution to the form factor at zero recoil is from the axial vector current $\bar{\psi}\gamma^\mu \gamma^5 \psi$:
\begin{align}
\langle V(p',\epsilon)|\bar{q}\gamma^\mu \gamma^5 Q|P(p)\rangle = \nonumber
& 2M_VA_0(q^2)\epsilon^*\cdot q/q^2 q^\mu \nonumber
 +(M_P+M_V)A_1(q^2)\Big[ \epsilon^{*\mu} - \epsilon^*\cdot q/q^2 q^\mu \Big]\nonumber \\
&-A_2(q^2)\frac{\epsilon^*\cdot q}{M_B+M_V}\Big[ p^\mu + p'^\mu - \frac{M_B^2-M_V^2}{q^2}q^\mu \Big]
\end{align}
This reduces to 
\begin{align}
\langle V(\vec{p}'=0,\epsilon)|&\bar{q}\gamma^j \gamma^5 Q|P(\vec{p}=0)\rangle = (M_P+M_V)A_1(q_{\omega = 1}^2)\epsilon^{*j}
\end{align}
for $j=1,2,3$ where also
\begin{align}
{F}(1) = h_{A_1}(1) = \frac{M_P+M_V}{(2M_P2M_V)^{1/2}}A_1(q_{\omega = 1}^2)
\end{align}

 In order to extract this quantity from lattice calculations we compute the set of Euclidean correlation functions
\begin{align}
C_{B2pt}(t)_{ij}&= \langle \mathcal{O}(t)_{Bi} \mathcal{O}^\dagger(0)_{Bj}     \rangle \nonumber\\
C^{\mu\nu}_{D^*2pt}(t)_{ij}&= \langle \mathcal{O}^\mu(t)_{D^*i} \mathcal{O}^{\dagger\nu}(0)_{D^*j}     \rangle \nonumber\\
C^{\mu\kappa}_{3pt}(T,t,0)_{ij}&=\langle \mathcal{O}^\mu(T)_{D^*i} \mathcal{J}^\kappa(t) \mathcal{O}^\dagger(0)_{Bj}     \rangle
\end{align}
where each operator $\mathcal{O}_i$ is projected onto zero spatial momentum by summing over spatial lattice points and the current $\mathcal{J}^\kappa$ is formed from appropriate combinations of NRQCD-HISQ lattice currents \cite{Matching2013}. $i$ and $j$ label different smearing functions.  These correlation functions can be expressed in terms of amplitudes and decaying exponentials by inserting a complete basis of states. Projecting onto zero momentum and setting $q = (M_B-M_{D^*},0,0,0)$ this gives
\begin{align}
C_{B2pt}(t)_{ij}&=  \sum_{n}\sum_{a=0,1} (-1)^{at} \frac{\sqrt{Z_{i,B_a}^{(n)} Z_{j,B_a}^{(n)}}}{2M_{B_a}^{(n)}}e^{-M_{B_a}^{(n)} t}\nonumber\\
C^{\mu\nu}_{D^*2pt}(t)_{ij}&= \sum_{n,s}\sum_{a=0,1}  (-1)^{at}  \frac{\sqrt{Z_{i,D^*_a}^{(n)} Z_{j,D^*_a}^{(n)}}}{2M_{D^*_a}^{(n)}}  \epsilon^{\mu}_s \epsilon^{\nu*}_s   \Big[ e^{-M_{D^*_a}^{(n)} t} + e^{-M_{D^*_a}^{(n)} (L_t - t)} \Big]   \nonumber\\
C^{\mu\kappa}_{3pt}(T,t,0)_{ij}&= \sum_{n,m,s} \sum_{a,b = 0,1}  \frac{\sqrt{Z_{j,B_b}^{(n)}}}{2M^{(n)}_{B_b}}  \frac{\sqrt{Z_{i,D^*_a}^{(m)}}}{2M^{(m)}_{D^*_a}} \epsilon^\mu_s \epsilon^{\kappa *}_s \nonumber \\ 
\times  (M^{(n)}_{B_b}+&M^{(m)}_{D^*_a})A^{(nm)}_{1,ba}(q^2) e^{-M^{(m)}_{D^*_a}(T-t) - M^{(n)}_{B_b}t} (-1)^{a(T-t) + bt}
\label{corrfuncts}
\end{align}
where $n$ and $m$ label excited states, $s$ is a spin index and we have included the dependence upon time doubled states, labeled by $a$ and $b$, that result from using staggered quarks \cite{HISQ}. It can be shown that the $Z^{1/2}$ and $A^{nm}_{1,ba}$ factors, given our choice of operators, are real \cite{timereversal}. The spin-sum rules for the polarisation vectors $\epsilon$ then require us to take $\nu = \mu = \kappa \neq 0$. We extract $h_{A_1}(1) = (M^{(0)}_{B_0}+M^{(0)}_{D^*_0})/({2M_{D^*_0}^{(0)} 2M_{B_0}^{(0)}})^{1/2} A^{(00)}_{1,00} $  using Bayesian fitting techniques \cite{Fitting}.

We work with an NRQCD-HISQ current which is perturbatively matched to the continuum current, the method for which is outlined in \cite{Matching2013}. To $\mathcal{O}({\Lambda}/{M_b})$ we only need
\begin{align}
{J^0_\text{latt}}^i(x) &= \bar{c} \gamma^i\gamma^5 Q ,
&{J^1_\text{latt}}^i(x) =-\frac{1}{2am_b} \bar{c} \gamma^i\gamma^5 \gamma  \cdot \Delta Q \nonumber
\end{align}
although we also compute matrix elements of
\begin{align}
{J^2_\text{latt}}^i(x) & = \frac{-1}{2am_b}\bar{c} \gamma \cdot \overleftarrow{\Delta} \gamma_0  \gamma^i\gamma^5  Q ,
&{J^3_\text{latt}}^i(x)  = \frac{-1}{2am_b}\bar{c}  \gamma^0\gamma^5  {\Delta}^i   Q. \nonumber 
\end{align}
Through $\mathcal{O}(\alpha_s, {\Lambda}/{M_b})$ the matched current is given by
\begin{align}
\mathcal{J}^i = Z( (1+\alpha_s(\eta-\tau)){J^0_\text{Latt}}^i + {J^1_\text{latt}}^i)\nonumber 
\end{align}
where $Z$ is the multiplicative factor coming from the tree level massive HISQ wavefunction renormalisation. In order to do this matching an appropriate scale for $\alpha_s$ must be chosen. We use $\alpha_V(q^* = 2/a)$.
\section{Chiral-Continuum Extrapolation and Results}
\label{sec:Conclusions}

\begin{table}[H]
\begin{tabular}{c c c c c c c c c c} 
\hline
          & $ \langle J^0_\text{Latt} \rangle$       & $ \langle  J^1_\text{Latt} \rangle$ &            Z         & $\eta$        & $\tau$         &$\alpha_s$& $ F(1) $\\            
\hline
1 & 0.952(13)& 0.00208(84)& 0.9930& $-$0.260(3)& 0.0163(1)& 0.346& 0.857(12)\\
2 & 0.965(19)& 0.0011(13)& 0.9933& $-$0.260(3)& 0.0165(1)& 0.344& 0.869(17)\\
3 & 0.923(12)& 0.00047(89)& 0.9930&$-$0.260(3)& 0.0165(1)& 0.343& 0.830(11)\\
4 & 0.901(20)& 0.00132(94)& 0.9972& $-$0.191(3)& 0.0216(1)& 0.311& 0.840(19)\\
5 & 0.914(17)& 0.00220(75)& 0.9974& $-$0.185(3)& 0.0221(1)& 0.308& 0.856(16)\\
6 & 0.943(12)& 0.00236(71)& 0.9974&$-$0.185(3)& 0.0221(1)& 0.307& 0.883(11)\\
7 & 0.873(11)& 0.00248(48)& 0.9994&$-$0.091(3)& 0.033(1)& 0.267& 0.846(11)\\
8 & 0.9049(66)& 0.00475(55)& 0.9994& $-$0.091(3)& 0.033(1)& 0.267& 0.8791(65)\\
 \hline
\end{tabular}
\caption{Fit results for the matrix elements of each lattice current together with the relevant matching parameters, computed in \cite{Matching2013}, used to compute $ F(1)$ on each set. }
\end{table}
\begin{figure}[H]
    \begin{center}
    \includegraphics[width=0.8\textwidth]{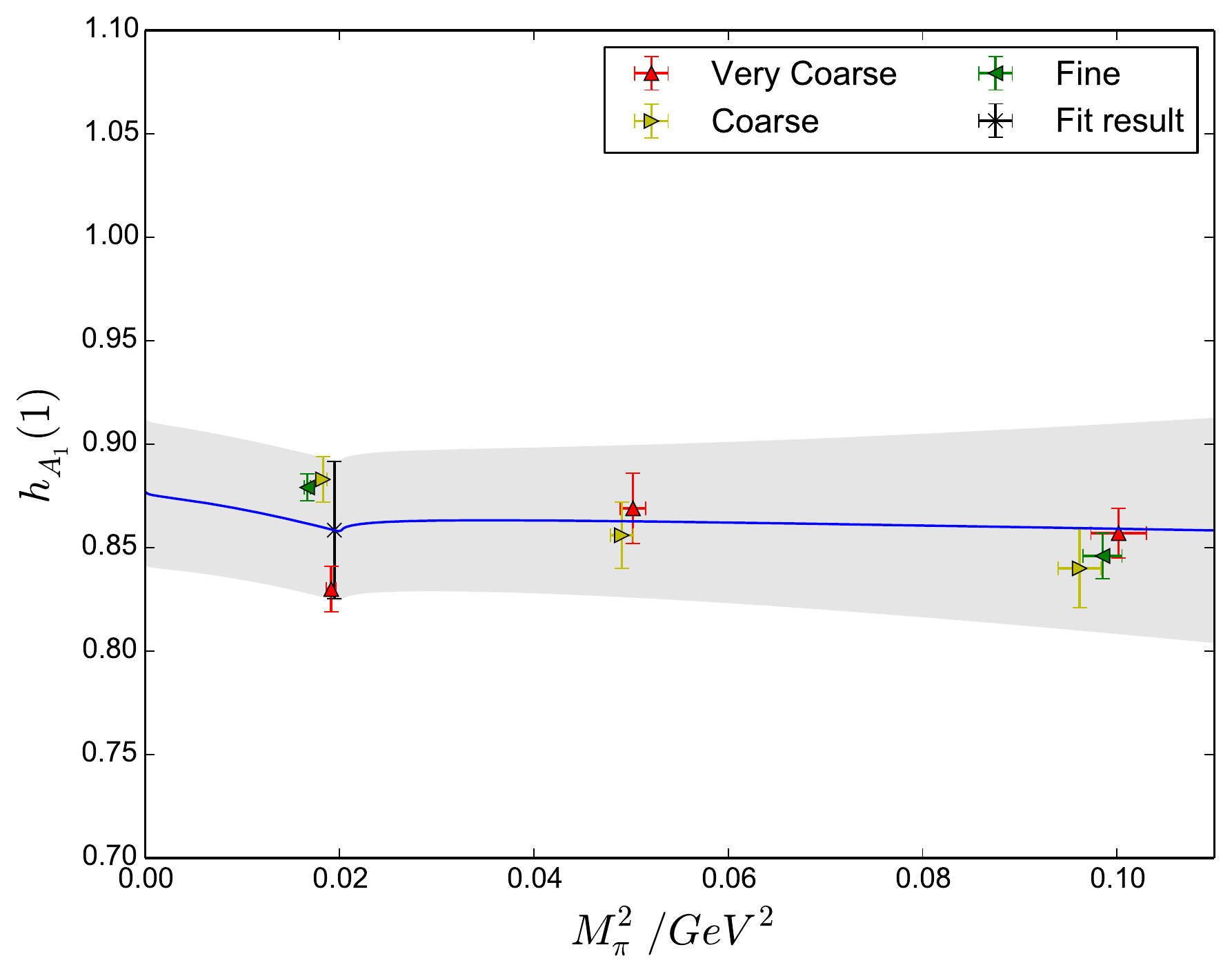}
    \end{center}
  \caption{A preliminary fit to our data using staggered chiral perturbation theory \cite{SCHIPT}. The grey band is the continuum chiral perturbation theory result extrapolated from our lattice data. It includes systematic errors coming from matching uncertainties and hence has a much larger error than any of the data points, which are only shown with their statistical error. We are working to investigate the apparent deviation of the very coarse result from the other two results at the physical pion mass.}
\end{figure}

Taking the result from this preliminary fit and combining it with the latest HFAG result \cite{HFAG}, $\eta_{EW} \mathcal{F}(1) |V_{cb}|= 35.81(11)(44)\times10^{-3}$, we find $V_{cb}=41.5(17)\times 10^{-3}$ where we have taken $\eta_{EW}=1.00662(16)$. Our result shows a slight tension with the determination by the Fermilab Lattice and MILC Collaborations \cite{FERMILABDSTAR} and is consistent with the inclusive result, with our error being dominated by the $\mathcal{O}(\alpha_s^2)$ matching uncertainty. It has also been suggested that this leading error could be constrained somewhat using semileptonic $B_c \rightarrow J/\psi$ decays and comparing to results from heavy-HISQ $b$ quarks on ultrafine lattices with $a=0.045\text{fm}$ and $m_ba<1$ \cite{AndrewTalk}.

\section*{Acknowledgements}
This work used the DiRAC Data Analytic system at the University of Cambridge,
operated by the University of Cambridge High Performance Computing Service on
behalf of the STFC DiRAC HPC Facility (www.dirac.ac.uk). This equipment was
funded by BIS National E-infrastructure capital grant (ST/K001590/1), STFC
capital grants ST/H008861/1 and ST/H00887X/1, and STFC DiRAC Operations grant
ST/K00333X/1. DiRAC is part of the National E-Infrastructure.

 \bibliography{Master}{}

\begin{thebibliography}{10}

\bibitem{REVIEW}
{ K.A. Olive and Particle Data Group},
\newblock Chin. Phys. C {\bf 38}, 090001 (2014).

\bibitem{FERMILABDSTAR}
Fermilab Lattice and MILC Collaborations, J.~A. Bailey {\em et~al.},
\newblock Phys. Rev. D {\bf 89}, 114504 (2014).

\bibitem{MILC1}
MILC Collaboration, A.~Bazavov {\em et~al.},
\newblock Phys. Rev. D {\bf 82}, 074501 (2010).

\bibitem{MILC2}
MILC Collaboration, A.~Bazavov {\em et~al.},
\newblock Phys. Rev. D {\bf 87}, 054505 (2013).

\bibitem{HISQ}
HPQCD Collaboration, E.~Follana {\em et~al.},
\newblock Phys. Rev. D {\bf 75}, 054502 (2007).

\bibitem{HFAG}
{Heavy Flavor Averaging Group},
\newblock aXiv 1412.7515 [hep-ex]  (2014).

\bibitem{ETAEW}
A.~Sirlin,
\newblock Nucl. Phys. B {\bf 196}, 83  (1982).

\bibitem{lightmassratio}
Fermilab Lattice and MILC Collaborations, A.~Bazavov {\em et~al.},
\newblock Phys. Rev. D {\bf 90}, 074509 (2014).

\bibitem{PhysMasses}
HPQCD Collaboration, B.~Chakraborty {\em et~al.},
\newblock Phys. Rev. D {\bf 91}, 054508 (2015).

\bibitem{GaugeAction}
HPQCD Collaboration, A.~Hart, G.~M. von Hippel, and R.~R. Horgan,
\newblock Phys. Rev. D {\bf 79}, 074008 (2009).

\bibitem{NRQCD}
G.~P. Lepage, L.~Magnea, C.~Nakhleh, U.~Magnea, and K.~Hornbostel,
\newblock Phys. Rev. D {\bf 46}, 4052 (1992).

\bibitem{latticespacing}
HPQCD Collaboration, R.~J. Dowdall {\em et~al.},
\newblock Phys. Rev. D {\bf 85}, 054509 (2012).

\bibitem{Matching2013}
C.~Monahan, J.~Shigemitsu, and R.~Horgan,
\newblock Phys. Rev. D {\bf 87}, 034017 (2013).

\bibitem{timereversal}
J.~J. Dudek, R.~G. Edwards, and D.~G. Richards,
\newblock Phys. Rev. D {\bf 73}, 074507 (2006).

\bibitem{Fitting}
{G.P. Lepage, B. Clark, C.T.H. Davies, K. Hornbostel, P.B. Mackenzie, C.
  Morningstar, H. Trottier},
\newblock arXiv hep-lat/0110175  (2001).

\bibitem{SCHIPT}
{J. Laiho, R.S. Van de Water},
\newblock Phys. Rev. D {\bf 73}, 054501 (2006).

\bibitem{AndrewTalk}
{A. Lytle},
\newblock {These proceedings}  ({2016}).

\end{thebibliography}
 \bibliographystyle{h-physrev5.bst}

\end{document}